\documentstyle[twoside,fleqn,espcrc2]{article}
\title{$\Delta S = 2$ decays of $B^-$ meson in MSSM and two 
Higgs doublet model}
\author{S. Fajfer\address{Physics Department, University of Ljubljana 
and J. Stefan Institute,  1000 Ljubljana, 
Slovenia} and 
P. Singer\address{Department of Physics, Technion - Israel Institute  of Technology, 
Haifa 32000, Israel}}
\begin{document}

\begin{abstract}

In view of the extreme smallness of $\Delta S = 2$ 
transitions of $B$ meson in the Standard Model, we consider their 
occurence in several extensions of it. Thus, we analyze the 
three - body $B^- \to K^- K^- \pi^+$ and two - body 
$B^- \to K^{*-} \bar K^{*0}$, 
$B^- \to K^-  \bar K^{0}$, $B^- \to K^{*-} \bar K^{0}$,   
 $B^- \to K^{-} \bar K^{*0}$ 
 decay modes both in the Standard Model and in the Minimal 
 Supersymmetric Model with and without ${\cal R}$ parity 
 conservation and in two Higgs doublet 
 models. All five modes are found to have a branching ratio 
 of the order of $10^{-13}$ in the Standard Model, while 
 the expected  
 branching ratio in the different extensions vary between 
 $10^{-9} - 10^{-6}$, for a given reasonable choice of 
 parameters. 
\end{abstract}
\maketitle

The rare B meson decays are very important in current  searches 
of 
physics beyond  Standard Model (SM) \cite{AM1}.
Recently, it has been suggested \cite{PAUL1,PAUL2,FS} to investigate effects of new 
physics possibly arising from $b \to ss \bar d$ or 
$b \to dd \bar s$  decays.  
As shown in  Ref. \cite{PAUL1}, the $b \to ss \bar d$ transition is
mediated in the standard model  by the box-diagram and its calculation
results in a branching ratio of nearly  $10^{-11}$, the exact value
depending on the relative unknown phase between t, c contributions  
in the box.
The authors of Refs. \cite{PAUL1,PAUL2} have 
calculated the $b \to ss \bar d$ transition in various 
extensions of the SM. It appears that for certain plausible
values of the parameters, this decay
may proceed with a branching ratio
of $10^{-8} - 10^{-7}$ in the minimal supersymmetric 
standard model (MSSM) 
and in two Higgs doublet models \cite{PAUL2}. 

Thus, decays related to the $b \to s s \bar d$ transition 
which was calculated to be very rare in the Standard Model, 
provide a good opportunity for investigating 
beyond the Standard Model physics. 
Moreover, when one considers
supersymmetric models with ${\cal R}$-parity violating couplings, 
it turned out that
the existing bounds on the involved couplings of the 
superpotential
did not provide any constraint on the $b \to s s \bar d$
mode \cite{PAUL1}. 
Recently, the OPAL collaboration \cite{OPAL} has set  
lower bounds on these 
couplings from the establishment of an upper limit for 
the $B^- \to K^- K^- \pi^+$ decay 
$BR(B^- \to K^- K^- \pi^+) \leq 8.8 \times 10^{-5
}$. 
Here we investigate 
the branching ratio of this decay mode in MSSM, with and without ${\cal R}$ parity 
and two Higgs doublet models as possible alternatives to the SM. Then we comment 
on another possibilty for the observation of the 
 $b \to s s \bar d$ transition: the two body decays of $B^-$. 
First, we proceed to describe the framework used in our analysis 
in which we concentrate on MSSM, with and without ${\cal R}$ parity 
and two Higgs doublet models.  \\
The minimal supersymmetric extension of the Standard Model  
leads to the following effective Hamiltonian describing the $b \to s s \bar d$ 
transition \cite{PAUL1,GGMS}
\begin{eqnarray}
{\cal H} &=& \tilde C_{MSSM}
(\bar s \gamma^\mu d_L ) (\bar s \gamma_\mu b_L), 
\label{1}
\end{eqnarray}
where we have denoted 
\begin{eqnarray}
\tilde C_{MSSM} & = &   -\frac{\alpha_s^2 \delta_{12}^{d*}\delta_{23}^d}{216 
m_{\tilde d}^2}[ 24 x f_6(x) + 66 \tilde f_6(x)] 
\label{2}
\end{eqnarray}
with $x = m_{\tilde g}^2 / m_{\tilde d}^2$, and the functions 
$f_6(x)$ and $ \tilde f_6(x)$ are given in \cite{GGMS}. 
The couplings 
$\delta_{ij}^{d}$ parametrize the mixing between 
the down-type left-handed squarks.
At the scale of
$b$ quark mass and by taking the existing upper limits  on 
$\delta_{ij}^{d}$ from \cite{GGMS} and \cite{PAUL1} 
the coupling $\tilde C_{MSSM}$ is estimated to be 
$|\tilde C_{MSSM}|\leq 1.2 \times 10^{-9}$ $ GeV^{-2}$ for an average 
squark mass $m_{\tilde d} = 500$ $GeV$ and $x =8$, which leads to 
an inclusive branching ratio for $b \to s s \bar d$ of 
$2 \times 10^{-7}$ \cite{PAUL1}.
The corresponding factor calculated in  SM, taking numerical values
from \cite{CASO} and  neglecting the CKM phases is estimated to be   
$|C_{SM}| \simeq 4 \times 10^{-12}$
\cite{PAUL1}.  

The authors of \cite{PAUL1}  have also investigated 
beyond MSSM cases  by including $R$- parity 
violating  interactions. 
 The part of the superpotential which is relevant here is 
 $W = \lambda_{ijk}^\prime L_i Q_j d_k$,
 where $i, j, k$ are indices for the families and $L, Q,d$ 
 are superfields for the
 lepton doublet, the quark doublet, and the down-type quark 
 singlet, respectively.
Following notations of \cite{CR} and \cite{PAUL1} 
the tree level effective Hamiltonian 
is 
\begin{eqnarray}
{\cal H} =- \sum_{ n} \frac{f_{QCD}}{m_{\tilde \nu_n}^2}
[\lambda^\prime_{n32} \lambda^{\prime *}_{n21} 
(\bar s_R b_L ) (\bar s_L d_R)&&\nonumber\\ 
+  \lambda^{\prime }_{n21} 
\lambda^{\prime *}_{n32} 
(\bar s_R d_L ) (\bar s_L b_R)]. &&
\label{4}
\end{eqnarray}
The QCD corrections were found to be important for this transition 
\cite{BAGGER}. For our purpose  it suffices to 
follow \cite{PAUL1} retaining the leading order QCD result 
$f_{QCD} \simeq 2$, for $m_{\tilde \nu}= 100$ $GeV$.

 Most recently an upper bound on the specific  combination of 
 couplings entering 
 (\ref{4}) has been obtained by OPAL from a search for the 
 $B^- \to K^- K^- \pi^+$ decay \cite{OPAL}  
 $\sum_{n} ( |\lambda^\prime_{n32} \lambda^{\prime *}_{n21}|^2 
+ |\lambda^{\prime }_{n21} 
\lambda^{\prime *}_{n32}|^2 )^{1/2}< 10^{-4}$. 
Here we take the order of magnitude, while the OPAL result is 
$5.9 \times10^{-4}$ based on a rough estimate 
$\Gamma (B^- \to K^- K^- \pi^+) \simeq 1/4~ $ 
$ \Gamma (b \to ss \bar d)$. 

The decay $b \to ss \bar d$ has been investigated using 
two Higgs doublet models (THDM) as well 
\cite{PAUL2}. These authors found that 
the charged Higgs box contribution 
in MSSM is negligible. On the other hand, THDM  
involving several neutral Higgses \cite{CS} could 
have a more sizable contribution to these modes. 
The part of the effective Hamiltonian relevant in our case is 
the tree diagram exchanging 
the neutral Higgs bosons $h$ (scalar) and $A$ (pseudoscalar) 
(see for details \cite{PAUL2,FS1}). 
\begin{eqnarray}
&&{\cal H}_{TH} = \frac{i}{2} \xi_{sb} \xi_{sd} (\frac{1}{m_h^2} 
(\bar s d ) (\bar s b)\nonumber\\
&& - \frac{1}{m_A^2}(\bar s \gamma_5 d ) 
(\bar s \gamma_5 b) ), 
\label{6}
\end{eqnarray}
The coupling $\xi_{ij}$ defined in \cite{CS} as a Yukawa 
coupling
of the FCNC transitions $d_i \leftrightarrow d_j$. 
In our estimation
we use the bound 
$|\xi_{sb} \xi_{sd}|/m_H^2 > 10^{-10}$ $ GeV^{-2}$, 
$H = h,A$,  which was obtained in  \cite{PAUL2} by using the 
$\Delta m_K$ limit on $ \xi_{bd}/m_H$ and assuming 
$| \xi_{sb}/m_H| >10^{-3}$. \\

We proceed now to study the effect of Hamiltonians (\ref{1}), 
(\ref{4}) 
on the various two body $\Delta S = 2$ decays of 
charged B - mesons. 
In order to calculate the matrix elements of the operators appearing in 
 the effective Hamiltonian, we use the factorization approximation 
 \cite{ALI,WSB1,WSB2}, which requires the knowledge of the 
 matrix elements of the current operators or the density operators. 
 Here we use the standard form factor representation 
 \cite{WSB1,ALI} of the 
 matrix elements described in detail in \cite{FS,FS1}. 

 For the  $F_1$ and $F_0$ form factors 
appearing in the decomposition 
of the matrix element of the weak current 
between two pseudoscalar states, one usually assumes pole
dominance \cite{WSB1,JURE}. 
For the vector and axial vector form factor, appearing in the decomposition 
of the  matrix element of the week current 
between the vector and  pseudoscalar states, we use again 
pole
dominance \cite{WSB1,JURE}. The relevant parametrs 
are taken from \cite{ALI,WSB2} $F_0^{BK}(0) = 0.38$, 
$A_0^{BK*}(0) = 0.32$. 
For the calculations of the density operators we use  derivatives of the vector or
axial-vector currents \cite{FS1}.  
In our numerical calculation we use  $f_K= 0.162$ $GeV$, $g_{K^*}= 0.196$  $GeV^2$ \cite{WSB2}. 
Now we turn to the analysis of the specific modes.\\

We denote ${\cal O} =
(\bar s \gamma^\mu(1 - \gamma_5) d )$ 
$ (\bar s \gamma_\mu (1 - \gamma_5)b)$, and then we use 
${\cal H} = C {\cal O}$ with $C$ being $1/4 \tilde C_{MSSM}$,   
$1/4 C_{SM}$ \cite{FS1}. 
Using factorization  and introducing  $s= (p_B - k_1)^2 $, $t = (p_B - k_2)^2 = $ 
and $ u = (p_B - p_\pi)^2$, one finds
for the $B^- \to K^- K^- \pi^+$ decay 
\begin{eqnarray}
\langle K^-(k_1) K^- (k_2) \pi^+ (p_\pi) | {\cal O} | B^-(p_B)\rangle =&&
\nonumber\\
F_1^{K\pi}(s) F_1^{BK}(s) [ m_B^2 + 2 m_K^2 - s - 2 t - && \nonumber\\
- \frac{m_K^2 - m_\pi^2}{s} (  m_B^2 - m_K^2 )]&& \nonumber\\
+ F_0^{K\pi}(s) F_0^{BK}(s)\frac{m_K^2 - m_\pi^2}{s} (  m_B^2 - m_K^2 ).
\label{15}
\end{eqnarray}
Within MSSM the branching ratio is found to be  
\begin{eqnarray}
BR(B^- \to K^- K^- \pi^+)_{MSSM}\leq 4.7 \times 10^{-9}, && 
\label{16}
\end{eqnarray}
while SM gives this rate to be $ 5.2 \times 10^{-14}$. 
The MSSM  which includes ${\cal R}$ parity breaking terms 
can occur in this decay. The matrix element of the 
operator ${\cal O}_{{\cal R}} = (\bar s (1 \pm \gamma_5) d )$ 
$ (\bar s (1 \mp \gamma_5) b)$ is found to be
\begin{eqnarray}
\langle K^-(k_1) K^- (k_2) \pi^+ (p_\pi) | {\cal O}_{{\cal R} }| B^-(p_B)\rangle =&&
\nonumber\\
F_0^{K\pi}(s) F_0^{BK}(s)\frac{ ( m_B^2 - m_K^2 )(m_K^2 - m_\pi^2)}{(m_s - m_d)
(m_b-m_s)}.
\label{R16}
 \end{eqnarray} 
Taking the values of the quark masses as in \cite{ALI}
$m_b = 4.88$ $GeV$, $m_s= 122$ $MeV$, $m_d = 7.6$ $MeV$ and using 
the bound given above, we estimate the upper limit of the branching ratio 
$ BR(B^- \to K^- K^- \pi^+)_{{\cal R}}$ $ \leq 1.8 \times 10^{-7}$. 
 This limit can be raised 
 to  $6\times 10^{-6}$ for the upper bound on the couplings of 
  $5.9 \times 10^{-4}$ given in \cite{OPAL}. 

The two Higgs doublet model, with the limit   
$|\xi_{sb} \xi_{sd}|/m_H^2 > 10^{-10}$ $ GeV^{-2}$, results in 
a branching ratio of the order $10^{-10}$. \\

The long distance effects (LD) are usually suppressed in the $B$ meson 
decays. However, in any search  of new physics one has to include their contributions also
\cite{FS}. In the case of  $B^- \to K^- K^- \pi^+$ decay,  we have 
analyzed two contributions \cite{FS}: (I) the box diagram, 
which is essentially the LD analog of the 
SD calculation in the standard model \cite{PAUL1} of 
the $b \to s s \bar d$ transition. (II)
 the contribution of virtual $"D^0"$
 and $"\pi^0"$ mesons, via the chain $B^- \to K^- "D^0"("\pi^0")$ 
 $\to K^- K^- \pi^+$. This contribution arises 
 as a sequence of two $\Delta S = 1$ transitions and may lead to 
 final $K^- K^- \pi^+$ state as well. It is therefore 
 necessary to have an estimate of its relevance vis - \`a - vis 
 the "direct" $\Delta S = 2$ transition.
The box diagrams contributes to the real and imaginary part of the amplitude
for the  $B^- \to K^- K^- \pi^+$ decay. 
We have found that the real part, for a reasonable value of the cut-off 
parameter $\Lambda \simeq 10 $ ${\rm GeV}$ results in the rate  
$8 \times 10^{-15}$, while the imaginary part of the amplitude leads to 
the rate  $6 \times 10^{-12}$. 
The largest nonresonant contribution of the   
$"D^0"$ or $"\pi^0"$ exchange comes from the $"D^0"$, giving 
a branching ratio smaller than 
$10^{-13}$.  
Therefore, we have  shown that the long - distance 
contributions to $B^- \to K^- K^- \pi^+$ 
in the SM are smaller or comparable to the short - distance box diagram, 
and have the branching ratio in  a $10^{-12} - 10^{-11}$ range. 
 This is a most welcome 
feature since it strengthens the suitabilty of the 
$B^- \to K^- K^- \pi^+$ decay as an ideal testing ground 
for physics beyond the standard model, as originally 
suggested in ref. \cite{PAUL1}.\\

We briefly discuss the two - body  $\Delta S = 2$ decays of $B^-$ 
meson. Although in principle 
two body decays would appear to be simpler to analyze, 
there is the complication of $K^0 - \bar K^0$ mixing. Hence one needs also 
a good estimate for the $b \to s \bar s d$ transitions as well. 
For the analysis of pseudoscalar meson decay to two vector mesons  
 $B^- \to K^{*-} \bar K^{*0}$ it is convenient to use helicity formalism 
 (see details in \cite{FS1}). 
Within MSSM model the branching ratio  
becomes $\leq 6.2 \times 10^{-9}$, while SM gives this rate to be  
$ 6.8 \times 10^{-14}$. 
The ${\cal R}$ - parity term described by the effective Hamiltonian (4) cannot be seen 
in this decay mode when factorization
approach is used,  since the density operator matrix element 
$\langle \bar K^{*0} | (\bar s d)| 0\rangle$ vanishes. The two Higgs doublet 
model also cannot be tested in this mode due to the same reason.\\
The use of factorization technique described 
above gives following results in the case of $B^- \to K^{*-} \bar K^0$ decay:
within MSSM the branching ratio is  
straightforwardly found to be 
$ BR( B^- \to K^{*-} \bar K^0)_{MSSM} \leq 1.6 \times 10^{-9}$ \cite{FS1}, 
which is comparable to the SM prediction of Ref.  \cite{ALI} for the 
$\Delta S = 0$ $ B^- \to K^{*-} K^0$ decay given as 
$ BR( B^- \to K^{*-} K^0) = 1  \times 10^{-9}$,  
$ 5 \times 10^{-9}$, $ 2 \times 10^{-9}$ obtained for the 
number of colours $N_c = 2$,  $N_c = 3$,  $N_c = \infty$, 
 respectively. The  SM calculation for the $\Delta S= 2$ transition 
leads to 
$ BR( B^- \to K^{*-} \bar K^0)_{SM} = 1.7 \times 10^{-14}$.  
The MSSM  which includes ${\cal R}$ parity breaking terms 
can occur in this decay. The 
estimation of the upper limit of the branching ratio gives 
$ BR( B^- \to K^{*-} \bar K^0)_{{\cal R}}\leq $ 
 $ 4.4 \times 10^{-8}$. This limit can be raised 
 to  $1.5\times 10^{-6}$ for the upper bound on the couplings of 
  $5.9 \times 10^{-4}$ given in \cite{OPAL}. 
The two Higgs doublet model (\ref{6}) gives for the limit $|\xi_{sb} \xi_{sd}|/m_H^2 > 10^{-10}$ $ GeV^{-2}$,
a branching ratio of the order $10^{-11}$. Due to
specific combination of the products of the scalar (pseudoscalar) 
densities this is the only decay which has nonvanishing amplitude 
within the factorization assumption.\\
For the  $B^- \to K^{-} \bar K^{*0}$ decay mode the branching ratio in MSSM 
is constrained to be  
$BR(B^- \to K^{-} \bar K^{*0})_{MSSM}$ $ \leq 5.9 \times 10^{-9}$ 
in comparison with SM result $6.5 \times 10^{-14}$.
The amplitude calculated in MSSM including ${\cal R}$ breaking and 
THDM vanishes, due to vanishing of the matrix element of the 
density operator for $\bar K^{*0}$
state. \\ 
The $B^- \to K^{-} \bar K^0$ decay offers the following:
the branching ratio for MSSM is found to be 
$BR(B^- \to K^- \bar K^0)_{MSSM} \leq 2.3 \times 10^{-9}$,
in comparison with the $2.5 \times 10^{-14}$ found in the SM. 
The matrix element of the $R$ parity breaking MSSM operator has nonvanishing 
value and the  constraint on the coupling constants $\leq  10^{-4}$ gives the bound 
$9.4 \times 10^{-8}$, while 
for the bound of $5.9 \times 10^{-4}$ for the coupling constants the rate 
$BR(B^- \to K^- \bar K^0)_{{\cal R}}$ can reach  
$3.3 \times 10^{-6}$. \\
One might wonder if the  long distance effects are important in  
two - body $\Delta S = 2$ $B^-$ decays. We have estimated the tree level 
contribution of the 
$D (D^*)$ which then goes into $K (K^*)$ via weak annihilation. 
We found that these  contributions give a branching ratio 
of the order $10^{-18}$ and therefore they can be safely 
neglected.
One might think that  the exchange of two intermediate states 
$D (D^*)$, $K (K^*)$  can introduce certain long distance 
contributions.
In decay $B \to "D"~ "K" \to "K"~"K"$ 
the first weak vertex arises from the decay  $B \to "D" ~"K"$ 
and the second weak 
vertex (see e.g. \cite{FS}) can be generally obtained
from the three body decays of $D\to KKK$. 
Therefore, we are quite confident to suggest that the long distance effects 
are not important in the two - body $\Delta S = 2$ $B$ decays.\\ 

We can summarize that in the $B^- \to K^- K^- \pi^+$ and two - body $B^- $ 
decays,  
the MSSM with the chosen set of parameters gives rates of the
order $10^{-9}- 10^{-8}$, while  the ${\cal R}$ 
parity breaking terms in the MSSM can be seen only in the  
$B^- \to K^- K^- \pi^+$,  $B^- \to K^{*-} \bar K^{0}$ and 
$B^- \to K^{-} \bar K^{0}$ decays. Let us turn now to the possibility of detecting these decay modes. 
The $B^- \to K^- K^- \pi^+$ seems to be the best candidate, 
since the other  modes we 
discussed,  have a $\bar K^0$ in the final states which 
complicates the possibilty of a detection because of $K^0$  
$- \bar K^0$ mixing \cite{FS1}.
These are the modes which as we mentioned are more difficult 
on the experimental side. 
The THDM model can give nonvanishing contribution  in the case of 
$B^- \to K^- K^- \pi^+$ and $B^- \to K^{*-} \bar K^{0}$ decays, 
with a rate too small to be seen. 
Thus, we conclude 
that the $B^- \to K^- K^- \pi^+$ decay is an ideal candidate to look for  
physics beyond  SM and that the 
$B^- \to K^{*-} \bar K^{*0}$,  $B^- \to K^{-} \bar K^{*0}$ decays offer this 
possibility also.


\begin{thebibliography}{10}
\bibitem{AM1}  For reviews, see A. Masiero, L. Silvestrini
in Proc. $2^{nd}$ Intern. Conf. on B Physics and CP Violation
(BCONF 97), Honolulu, HI 1997, T. E. Browder,
F. A. Harris, S. Pakvasa eds (World Scientific, Singapore 1998),
p.172.  Y. Grossman, Y. Nir, R. Rattazzi in "Heavy Flavours"
 ($2^{nd}$ Edition), A. J. Buras and M. Lindner eds, p. 755
  (World Scientific, Singapore 1998).
\bibitem{PAUL1} K. Huitu, C. -D.L\" u , P. Singer,
D. -X. Zhang, Phys. Rev. Lett. 81 (1998) 4313.
\bibitem{PAUL2} K. Huitu, C. -D.L\" u , P. Singer,
D. -X. Zhang,Phys. Lett. B 445 (1999) 394.
\bibitem{FS} S. Fajfer and P. Singer, Phys. Lett. B 478 (2000) 185.   
\bibitem{OPAL} G. Abbiendi et al., OPAL Collaboration, Phys. 
Lett. B  476 (2000)  233.
\bibitem{GGMS} F. Gabbiani, E. Gabrielli, A. 
 Masiero, L. Silvestrini, Nucl. Phys. B 477 (1996) 321. 
\bibitem{CASO} Particle Data Group, D. E. Groom et. al, Europ. Phys.
Journal C  15 (2000) 1.
\bibitem{CR} D. Choudhury and P. Roy, Phys. Lett. B 378  (1996) 153.
\bibitem{BAGGER} J. L. A. Bagger, K.T. Matchev, and R.J. Zhang, 
Phys. Lett. B 412 (1997) 77.  
\bibitem{CS} T. P. Cheng, M. Sher, Phys. Rev. D 35 (1987) 3484; 
 M. Sher, Y. Yuan, Phys. Rev. D 41 (1991) 1461.
  \bibitem{ALI} A. Ali, G. Kramer and C. D.  L\"u, Phys. Rev. D 
58 (1998) 094009. 
 \bibitem{WSB1} M. Wirbel, B. Stech and M. Bauer,
Z. Phys. C  29 (1985) 637.
\bibitem{WSB2} M. Bauer, B. Stech, M. Wirbel, Z. Phys. C 34 (1987)  103; 
F. Buccella, M. Forte, G. Miele, G. Ricciardi,
Z. Phys. C 48 (1990) 47; P. Lichard, Phys. Rev. D 55 (1997)  
5385. 
\bibitem{JURE} S. Fajfer, J. Zupan, Int. J. Mod. Phys. A  14 (1999)  
4161. 
\bibitem{FS1} S. Fajfer and P. Singer, hep-ph/0007132.



\end{thebibliography}
\end{document}